\documentclass[final,5p,times,twocolumn]{elsarticle}

\usepackage{graphicx}
\usepackage{bm}
\usepackage{txfonts}
\usepackage{booktabs}
\usepackage{color}
\usepackage{multirow}
\usepackage{amssymb}
\usepackage{amsthm}

\newcommand{\Hc}{\ensuremath{H_{\rm c}}}

\newcommand{\Tc}{\ensuremath{T_{\rm c}}}
\newcommand{\HAC}{\ensuremath{H_{\rm AC}}}
\newcommand{\HDC}{\ensuremath{H_{\rm DC}}}
\newcommand{\cAC}{\ensuremath{\chi_{_{\rm AC}}}}
\newcommand{\kGL}{\ensuremath{\kappa_{\rm GL}}}
\newcommand{\mOc}{\ensuremath{\muup\Omega{\rm cm}}}

\journal{Physica C}

\begin{document}

\begin{frontmatter}

\title{AC susceptibility study of superconducting aluminium-doped silicon carbide}

\author[Kyoto]{M.~Kriener\corref{cor1}}
\ead{mkriener@scphys.kyoto-u.ac.jp}
\cortext[cor1]{Corresponding author}
\author[Tokyo]{T.~Muranaka}
\author[Tokyo]{J.~Akimitsu}
\author[Kyoto]{Y.~Maeno}

\address[Kyoto]{Department of Physics, Graduate School of Science, Kyoto University, Kyoto 606-8502, Japan}

\address[Tokyo]{Department of Physics and Mathematics, Aoyama-Gakuin University, Sagamihara, Kanagawa 229-8558, Japan}

\begin{abstract}
In 2007, type-I superconductivity in heavily boron-doped silicon carbide was discovered. The question arose, if it is possible to achieve a superconducting phase by introducing dopants different from boron. Recently, aluminum-doped silicon carbide was successfully found to superconduct by means of resistivity and DC magnetization measurements \cite{muranaka09a}. In contrast to boron-doped silicon carbide, the aluminum doped system is treated as a type-II superconductor because of the absence of an hysteresis in data measured upon decreasing and increasing temperature in finite magnetic fields. In this paper, results of a recent AC susceptibility study on aluminum-doped silicon carbide are presented. In higher applied DC magnetic fields and at low temperatures, a weak indication of supercooling with a width of a few mK is found. This supports the conclusion that aluminum-doped silicon carbide is located near to the border between type-I and type-II superconductivity, as pointed out in a recent theoretical work, too.

\end{abstract}


\end{frontmatter}

\section{Introduction}
Silicon carbide is well-known for its huge number of crystal modifications, all of them breaking inversion symmetry. Some publications consider more than 200 of them with either cubic (''C''), hexagonal (''H''), or rombohedral (''R'') symmetry of the unit cell, see e.\,g.\ \cite{casady96a,herrero09a}. They are usually referred to as $m$C-SiC, $m$H-SiC, and $m$R-SiC, respectively. The variable $m$ gives the number of Si\,--\,C bilayers consisting of a C and a Si layer stacking in the unit cell. Further information and sketches of the unit cells of 3C-SiC and 6H-SiC can be found in Ref.\,\cite{kriener08a} (see Fig.\,1 and Table\,1 therein). In 2007, superconductivity was found in silicon carbide upon heavy boron-doping (SiC:B) at boron concentrations of about $1-2\times 10^{21}$\,cm$^{-3}$ \cite{ren07a}. 
The superconducting transition temperature \Tc\ in zero field is about 1.45\,K and the upper critical field strength $\sim 115$\,Oe \cite{ren07a}. Moreover, a strong supercooling effect indicated by a large hysteresis in temperature- (field-) dependent AC susceptibility measurements upon decreasing and increasing temperature (field down and up sweep) in finite magnetic DC fields was observed. This led to the conclusion that SiC:B is a type-I superconductor. An estimate of the Ginzburg-Landau (GL) parameter \kGL\ from the charge-carrier concentration, the Sommerfeld parameter of the normal-state specific heat, and the critical temperature yielded $\kGL = 0.35$ \cite{kriener08a} in agreement with the type-I superconductivity in this system which needs $\kGL<1/\sqrt{2}$. The next step was to prepare SiC crystals with dopants other than boron. Recently, aluminum doping was found to induce superconductivity in 3C-SiC (SiC:Al) at a charge-carrier concentration of about $n\approx 0.7\times 10^{21}$\,cm$^{-3}$ \cite{muranaka09a}. Suprisingly, the zero-field critical temperature remains almost unchanged 1.5\,K. 

In this paper, an AC susceptibility study on SiC:Al is presented.
The preparation of the aluminium-doped SiC samples was done in a similar way as for SiC:B, see \cite{ren07a}. 
The sample used is a polycrystalline sample and not single phase consisting mainly of 3C-SiC and unreacted Al and Si phase fractions. An analysis of powder x-ray diffraction measurements suggests that Al substitutes at the Si site in SiC. In resistivity data, a sharp transition into a superconducting ground state at $\Tc=1.5$\,K is observed. The residual resistivity $\rho_0$ at \Tc\ is about 0.75\,\mOc. Above \Tc, the system features a metallic-like temperature dependence with a positive slope of d$\rho/$d$T$ in the whole temperature range up to room temperature and a residual resistivity ratio ${\rm RRR}=\rho({\rm 300\,K})/\rho_0$ of about 5.
In temperature-dependent resistivity measurements at constant DC magnetic fields as well as field-dependent measurements at constant temperatures, no indication for a supercooling behavior was found in contrast to SiC:B leading to the conclusion that the superconductivity in SiC:Al is of type-II rather than type-I. 

\section{Results and discussion}
AC susceptibility measurements were performed using a mutual-inductance method in a commercial $^3$He refrigerator (Oxford Instruments, Heliox) inserted into a standard superconducting magnet. A small AC field of $\HAC = 0.01$\,Oe-rms at a frequency of 3011\,Hz was used in zero and several finite DC magnetic fields \HDC\ parallel to \HAC\ as well as at constant temperatures upon sweeping the external field. 

\begin{figure}
\centering
\includegraphics[width=7.5cm,clip]{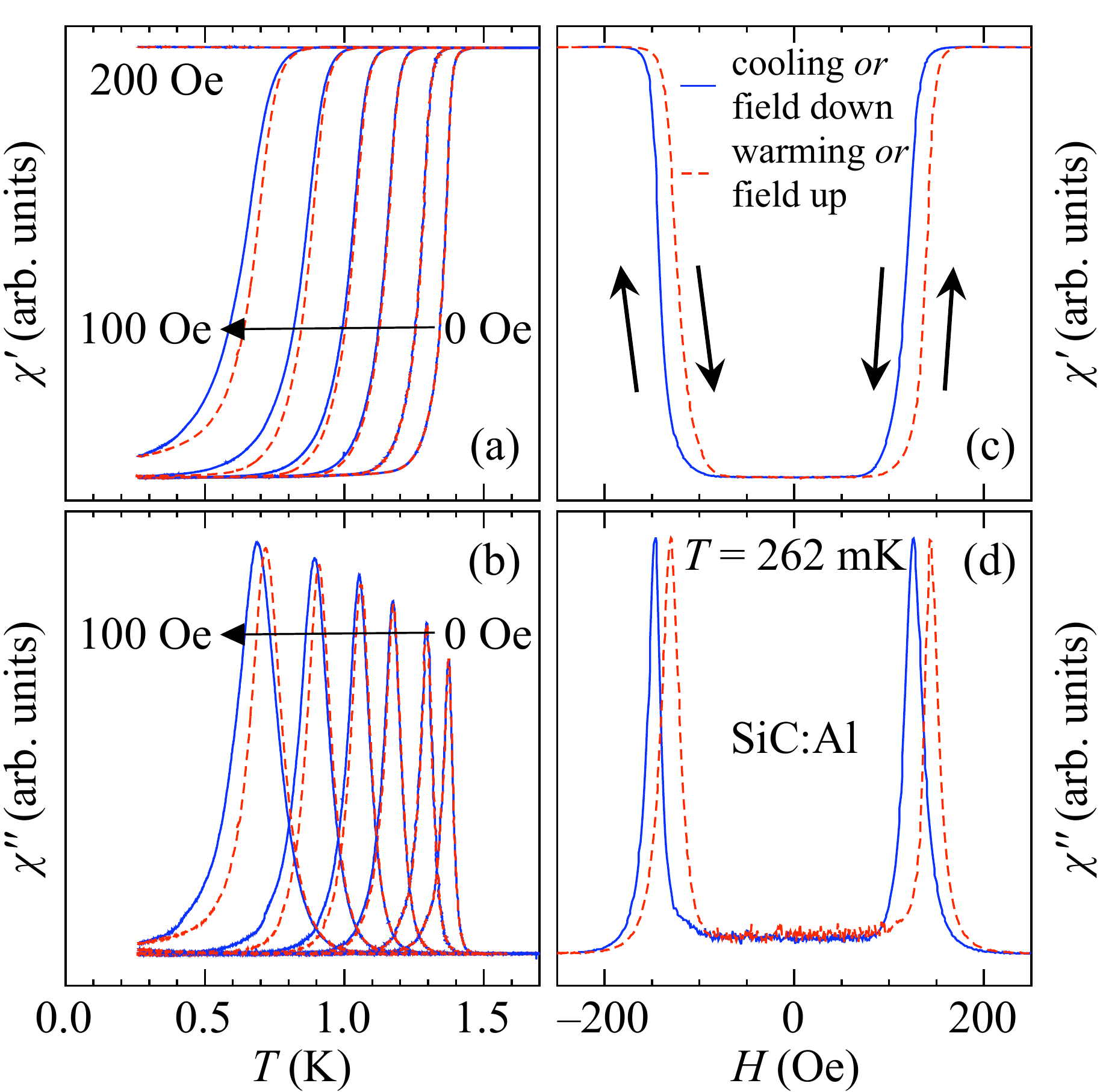}
\caption[]{(Color online) Temperature (panels (a) and (b)) and field dependence (panels (c) and (d)) of the AC susceptibility $\cAC=\chi'+i\chi''$ of SiC:Al. The upper two panels show the real part (shielding fraction) and the lower panels the imaginary part (energy dissipation) of \cAC. The solid (dashed) lines are measured upon decreasing (increasing) temperature / field. In panels (a) and (b) the arrows denote the direction of increasing DC magnetic fields. In panel (c) the arrows denote the sweep direction of the external magnetic field.} \label{AC1_Si5C4Al03}
\end{figure}
Fig.\,\ref{AC1_Si5C4Al03} summarizes AC susceptiblity $\cAC=\chi'+i\chi''$ data of SiC:Al. The temperature dependence in zero and several finite DC magnetic fields $\HDC=0$\,--\,200\,Oe of the real part or shielding fraction $\chi'$ is shown in Fig.\,\ref{AC1_Si5C4Al03}\,(a), in panel (b) the imaginary part or energy dissipation $\chi''$ is given. 
The temperature dependence of the in-field susceptibility data was taken as follows: The external DC magnetic field \HDC\ was set above \Tc. Then the temperature was reduced down to approx.\ 300\,mK and subsequently increased above \Tc. Panels (c) ($\chi'$) and (d) ($\chi''$) show field-dependent data taken after the sample was zero-field cooled to $T=262$\,mK. The arrows in panel (c) indicate the sweep direction of the external magnetic field. The resulting curves are labeled as ''cooling'' or ''field down'' (solid blue lines) and ''warming'' or ''field up'' (dashed red lines) in Fig.\,\ref{AC1_Si5C4Al03}. Here, \Tc\ is defined as the temperature at which the absolute value of $\chi'$ decreased by 1\,\% of the total difference in the signal between the normal and the superconducting state.

In zero field, a single sharp transition with a \Tc\ of 1.45\,K is observed, in good agreement with the resistivity data. In finite magnetic fields, the transition slightly broadens with increasing \HDC. In contrast to the resistivity data reported in Ref.\,\cite{muranaka09a}, the AC susceptibility data reveals a hysteresis at higher DC magnetic fields albeit this effect is rather tiny. The hysteresis increases with increasing the external field strength / lowering the temperature as indicated by the difference between the corresponding solid (cooling) and dashed (subsequent warming run) lines in Figs.\,\ref{AC1_Si5C4Al03}\,(a) and (b) or between the respective lines (dashed $\leftrightarrow$ field sweep up, solid $\leftrightarrow$ field sweep down) in Figs.\,\ref{AC1_Si5C4Al03}\,(c) and (d). The difference of the onset temperature / field of the superconducting transition amounts to $\Delta \Tc \approx 8$\,mK at $\HDC = 100$\,Oe and $\Delta \Hc \approx 13$\,Oe at $T = 262$\,mK. 

In a recent theoretical study \cite{yanase09a}, the authors discuss the superconductivity in boron-doped diamond C:B and boron- and aluminum-doped SiC. They propose a phase diagram depending on the origin of the superconductivity, i.\,e., whether it emerges from host bands or from impurity bands, arguing that the type-II superconductor C:B is in the crossover region between host-band and impurity-band superconductivity, whereas type-I SiC:B is clearly allocated in the host-band regime. Then type-II SiC:Al would be closer to C:B in their phase diagram than to SiC:B. Such allocations originate from the difference in the doping site, i.\,e., the C site in case of boron and the Si site in case of aluminum doping. However, the observation of a fairly small difference between subsequent cooling (field down) and warming runs (field up sweeps) at higher magnetic fields / low temperatures can also be explained as follows. Assuming that pure single-phase SiC:Al is a type-I superconductor, this ''clean' feature could be ''hidden away'' by the disorder of the multiphase sample used and hence SiC:Al appears as a ''dirty'' type-II superconductor. This could lead to a sufficient suppression of supercooling in SiC:Al, see Refs.\,\cite{muranaka09a} and \cite{blase09a}.

\section{Summary}
The present AC susceptibility study is complementary to the resistivity and DC magnetization study on aluminum-doped silicon carbide in Ref.\,\cite{muranaka09a}. The high sensitivity of the AC susceptibility at the onset of a superconducting transition reveals a rather tiny difference in the onset temperatures of superconductivity upon cooling and warming in higher external magnetic fields. This hysteresis is attributable to supercooling. One may speculate that a ''clean'' single crystalline sample SiC:Al would exhibit a larger hysteresis. Therefore, further improvements of the sample quality are required.

\section*{Acknowledgments} 
This work was supported by a Grants-in-Aid for the Global COE ''The Next Generation of Physics, Spun from Universality and Emergence'' from the Ministry of Education, Culture, Sports, Science, and Technology (MEXT) of Japan, and by the 21st century COE program ''High-Tech Research Center'' Project for Private Universities: matching fund subsidy from MEXT. It has also been supported by Grants-in-Aid for Scientific Research from MEXT and from the Japan Society for the Promotion of Science (JSPS). TM is supported by Grant-in-Aid for Young Scientists (B) (No. 20740202) from MEXT and MK is supported as a JSPS Postdoctoral Research Fellow.

\end{document}